\documentclass[onecolumn,amsmath,amssymb]{revtex4}

\usepackage{graphicx}
\usepackage{graphics}
\usepackage{subfigure}
\usepackage{dcolumn}
\usepackage{bm}
\usepackage{txfonts}

\begin{document}

\title{Working with Large Language Models to Enhance Messaging Effectiveness for Vaccine Confidence}
\author{Lucinda Gullison$^{1}$}
\author{Feng Fu$^{1,2}$}
\email{fufeng@gmail.com}

\affiliation{ $^1$Department of Mathematics, Dartmouth College, Hanover, NH 03755, USA\\
$^2$Department of Biomedical Data Science, Geisel School of Medicine at Dartmouth, Lebanon, NH 03756, USA}

\date{\today}

\begin{abstract}
Vaccine hesitancy and misinformation are significant barriers to achieving widespread vaccination coverage. Smaller public health departments may not have expertise or resources for effective vaccine messaging. This paper explores the potential of ChatGPT-augmented messaging to promote confidence in vaccination uptake. We conducted a survey where participants chose between a pair of vaccination messages and assessed which was more persuasive and to what extent. In each pair of messages, there was was one original message and one ChatGPT-augmented message. At the end of the survey, participants were told that half of the messages were generated by ChatGPT. Participants were asked for a quantitative and qualitative response about how knowledge of a message's ChatGPT status changed their impression of the messages. Overall, ChatGPT messages were rated slightly higher than the original messages. ChatGPT messages generally scored higher when they were longer. Respondents did not express major concerns about ChatGPT messaging, nor was there a significant relationship between views on ChatGPT and message ratings. Finally, there was a correlation between whether the message was positioned first or second in the question and the message’s score. Overall, our results point to the potential of ChatGPT to improve vaccination messaging with potential for future research to clarify this human-AI collaborative relationship. 
\end{abstract}

\maketitle

\section{Introduction}

Public health campaigns are of the utmost importance in influencing individual behavior to achieve public health goals. Mass media campaigns in particular have been shown to have a substantial effect on behaviors related to health and risk~\cite{wakefield2010use}. As social media has become a dominant platform through which individuals receive information, public health agencies have utilized platforms like Twitter/X (hereafter referred to as Twitter for simplicity) to conduct media campaigns, with varying degrees of success. In particular, local public health agencies have taken to Twitter to combat vaccine hesitancy. Agencies that already have a large following tend to receive higher engagement, although word choice can also increase a message’s effectiveness~\cite{mayberry2023analyzing}. 

The COVID-19 pandemic has brought about unprecedented challenges to public health worldwide. In response, pharmaceutical interventions like vaccination have been crucial for controlling the spread of the disease and minimizing its impact on individuals and society. However, vaccine hesitancy and misinformation have also increased during COVID-19 and have presented a significant barrier to achieving widespread vaccination coverage~\cite{wiysonge2022vaccine}. Research shows that COVID-19 vaccination messaging has had mixed results. Messaging has been proven to increase uptake in many instances~\cite{rodriguez2023effect}.Therefore, it is crucial for public health to understand how vaccine messaging can be made more effective.

Smaller resource-limited departments may struggle with costs~\cite{athey2023digital} associated with producing effective messaging, even if the financial and health benefits of successful messaging might eventually outweigh the initial costs. As many smaller local organizations and departments do not have public relations teams or the resources to outsource advertising, cost-effective mechanisms for improving and generating vaccine messaging are needed.

To address this challenge, innovative approaches are needed to effectively communicate the benefits and importance of vaccination to the public. One such approach is the use of artificial intelligence (AI) and natural language generation (NLG) technologies, which can generate tailored and personalized messaging that resonates with different audiences~\cite{karinshak2023working,sohail2024chatgpt,lim2023artificial,ayers2023comparing,breum2024persuasive,carrasco2024large}.

In recent years, the development of large language models (LLM), such as GPT-3, has enabled the generation of high-quality and contextually relevant text across a wide range of applications via human-AI collaboration~\cite{vaccaro2024combinations}. ChatGPT, a language model based on the GPT-3 architecture, has shown promising results in generating engaging and informative messaging for various domains, including healthcare. ChatGPT also poses itself as a more accessible messaging tool than traditional media campaigns or public relations strategies. Recent work compares and evaluates the persuasiveness of different LLMs~\cite{matz2024potential}, finding ChatGPT is among one of the best performing models~\cite{hackenburg2024scaling}. 

That being said, there are concerns involving the use of ChatGPT in public health messaging. One concern is ChatGPT’s tendency to hallucinate \cite{alkaissi2023artificial}, or make up information that is untrue. This is a significant concern given that misinformation related to vaccination is already so prevalent on the internet \cite{garett2021online}. Another concern is around the perception of ChatGPT and other artificial intelligence technologies, which varies greatly but is oftentimes negative or fearful \cite{cugurullo2023fear}. There is potential for the efficacy of ChatGPT generated messaging to be undermined if people are aware and fearful of the technology’s role in creating the message.

In this paper, we explore the potential of ChatGPT-augmented messaging for promoting confidence in vaccination uptake and countering vaccine hesitancy. Specifically, we investigate the effectiveness of ChatGPT-augmented messaging compared to original messaging currently in circulation on the internet. We refer to the former messages as ChatGPT-augmented, as they are not completely generated by ChatGPT. Instead we work with ChatGPT by including original messages in the prompt. It is important to note that the use of ChatGPT is intended to ``augment'' the messaging, but whether this is achieved as desired depends on the public's perception. To this end, we use the crowdsourcing platform MTurk to help evaluate and compare ChatGPT-augmented messages with the original ones, and examine whether they are indeed perceived as more convincing than their original counterparts.

Our study builds on previous studies~\cite{karinshak2023working,salvi2024conversational,ayers2023comparing,sohail2024chatgpt} and provides insights into perceived persuasiveness towards messaging generated by human-AI collaboration~\cite{chen2024large}, rather than solely by LLM~\cite{matz2024potential,sohail2024chatgpt}, in a specific domain related to vaccination. Our results show that ChatGPT-augmented messages were indeed rated slightly higher than the original messages. ChatGPT-augmented messages generally scored higher when they were longer. Overall, our study contributes to the growing body of research on the potential for harnessing AI technologies to address pressing public health issues and improve healthcare communication.

\section{Results}

The comparison survey, implemented using Qualtrics, was presented as a study on vaccination messaging without disclosing the use of ChatGPT (see details in Data and Methods section). Participants were randomly assigned to view message pairs labeled A and B—one original and one augmented by ChatGPT—and asked to select the more convincing message and rate its persuasiveness on a scale of 1 to 5 (-1 to -5 meaning less convincing, 1 to 5 meaning more convincing). The questions comprised four demographic questions (age, race, gender, education level), six A/B message comparison questions, one comprehension test, and two exit questions at the end of the survey about their opinions regarding the use of ChatGPT and AI. We administered the survey to 200 participants recruited via Amazon Mechanical Turk (MTurk), and subjects were compensated with a cash bonus upon completion. The sequence of A and B was randomized to ensure unbiased comparisons. To ensure data reliability, measures such as preventing multiple responses, filtering duplicate IP addresses, and excluding responses that failed the comprehension test were implemented, resulting in 138 valid responses for the final analysis.

\begin{figure}[htp]
\centering
\includegraphics[width=8cm]{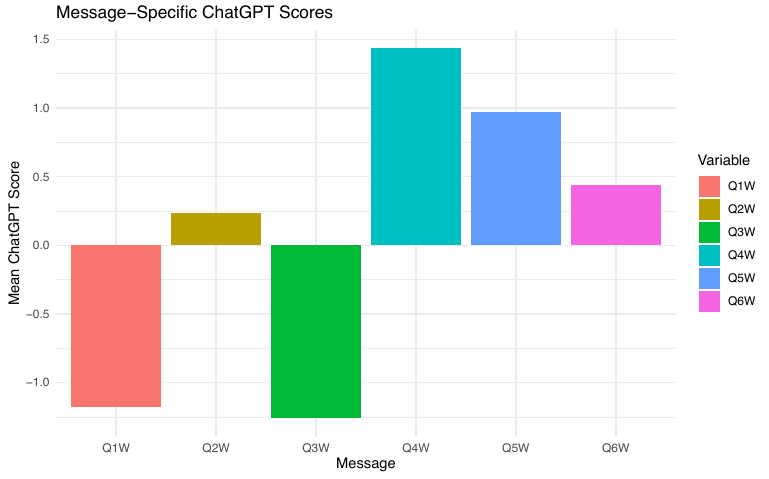}
\caption{Overall subjects' ratings favored ChatGPT-augmented messages. Out of the six ChatGPT-augmented messages, four were consistently rated higher. Subjects' responses were typically bimodal, with an overall average slightly positive, suggesting that ChatGPT-augmented messages are generally perceived more favorable by subjects.}
\label{fig1}
\end{figure}

The majority of the six messages had average scores that indicated a preference for the ChatGPT augmented message (Fig.~\ref{fig1}). Fig.~\ref{fig1} shows the average subjects' ratings of each of the six ChatGPT-augmented messages relative to the original ones, based on their convincingness and persuasion. Of the six sets of messages, four ChatGPT-augmented messages were preferred, while two original messages were preferred. We ran an ANOVA test across messages which yielded a p-value of $6.68e-10$. Therefore, there is a statistically significant difference among the mean scores of the six messages in our study.

Further analysis of the frequency distribution of mean scores of each respondent regarding ChatGPT-augmented messages shows a bimodal distribution with modes of -4 and 4. This result suggests that respondents either rated the ChatGPT-augmented messages very highly or very poorly. This observation may be attributable to the nature of how similar the messages were to each other, or how polarized/bimodal the responses were. Further studies might be needed to find a statistically significant relationship between whether a message is truly augmented by Chat-GPT and its average subject's rating. 

While the overall score of ChatGPT-augmented messages was not statistically significant enough to be positive ($n = 138$, mean: 0.1074879,  t-test: t = 0.73401, p-value = 0.4642), the high ratings of certain messages point to ChatGPT-augmented messages being highly effective in certain cases. In what follows, we investigate the potential explanations for the differences between ChatGPT-augmented messages in their subject's ratings.

\begin{figure}[htp]
\centering
\includegraphics[width=8cm]{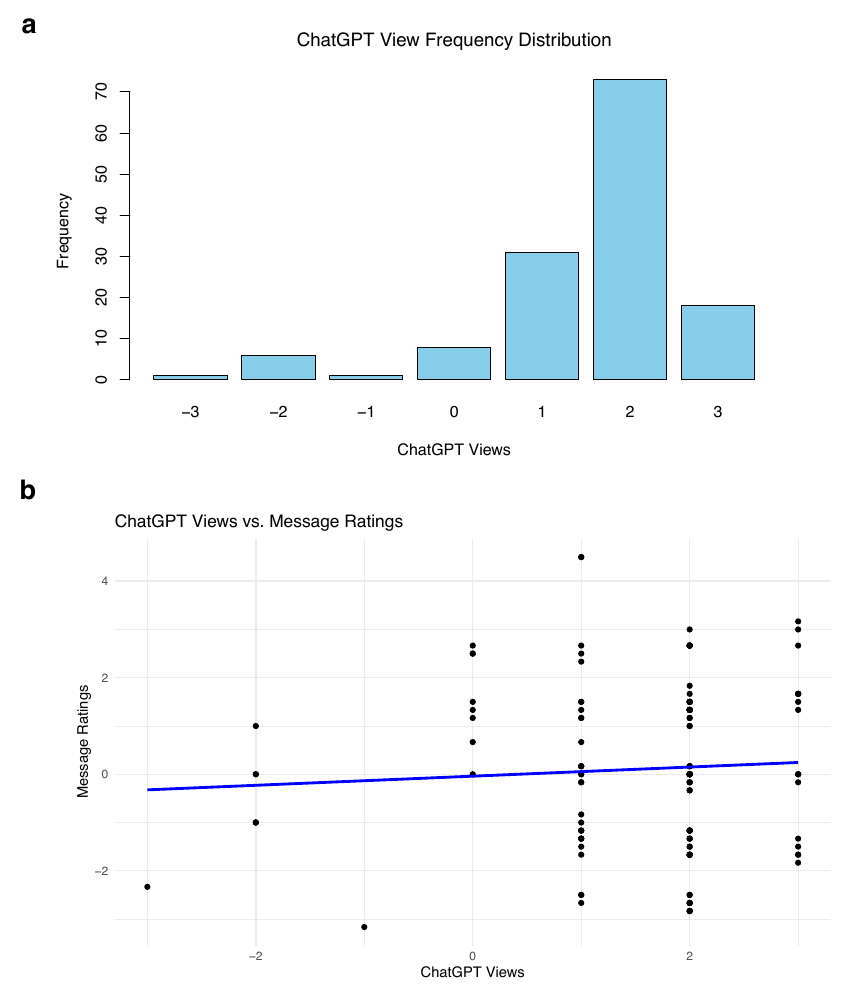}
\caption{Subjects' positive views on ChatGPT have little impact on their message ratings. The histogram in panel (a) shows that subjects generally hold positive views of ChatGPT with a few exceptions ($n=138$, mean  1.557971, one sample t-test p-value $<2.2e-16$ ). Scatter plot of subjects' ChatGPT view and their message ratings in panel (b), along with correlational analysis ($n = 138$, correlation coeff $r =  0.06366351$, p-value:  0.458192), suggests that subjects' positive views on ChatGPT do not lead them to consistently rate ChatGPT-augmented message higher.}
\label{fig2}
\end{figure}

Respondents generally had positive views on ChatGPT-augmented messaging (Fig.~\ref{fig2}a). The mean view of ChatGPT messaging was 1.557971 on a scale from -3 to 3, with 3 being the most positive and -3 being the most negative. 
122 responses were positive values, with 8 negative responses, and 8 responses of zero. We ran a one sample t-test with an alternative hypothesis that the true mean was not equal to 0. The t-test yielded a p-value < 2.2e-16, so at an alpha level of 0.05, we reject the null hypothesis and find a statistically significant result that respondents held overall positive attitudes toward ChatGPT. 

Of the respondents who left qualitative answers, the most frequently used word was ``good.'' That said, a minority of respondents viewed ChatGPT messages very negatively, and their qualitative responses indicated fears over ChatGPT replacing workers. Some described the messages as ``a bit over the top in conveying their message''. Another respondent expressed concerns about intellectual property issues with ChatGPT. These results suggest that there is a small subset of the population that is uncomfortable and distrustful of ChatGPT messaging. If the audience of a potential message ends up being part of this demographic opposed to ChatGPT generated messaging, that may be an issue for those looking to use ChatGPT. However, the overall positive views on ChatGPT suggest that this should not be a major concern when targeting the general public.

Next, we examined the relationship between views on ChatGPT and the scores that respondents gave the messages (Fig.~\ref{fig2}b). There was a small positive correlation between a participant’s mean message rating and their view on ChatGPT. The small positive correlation could be explained by people who are more approving of ChatGPT being more receptive to the style of writing that it produces.

The correlation coefficient for this relationship was 0.06366351 with a P-value of 0.458192. At an alpha level of 0.05, this result is not statistically significant, which suggests that there is little to no correlation between how respondents feel about ChatGPT and how they rate ChatGPT generated messages when they do not know whether the message was generated by ChatGPT or not. These results may differ if participants were told ahead of time that some of the messages were generated by ChatGPT, or more specifically, which ones were generated by ChatGPT.

\begin{figure}[htp]
\centering
\includegraphics[width=8cm]{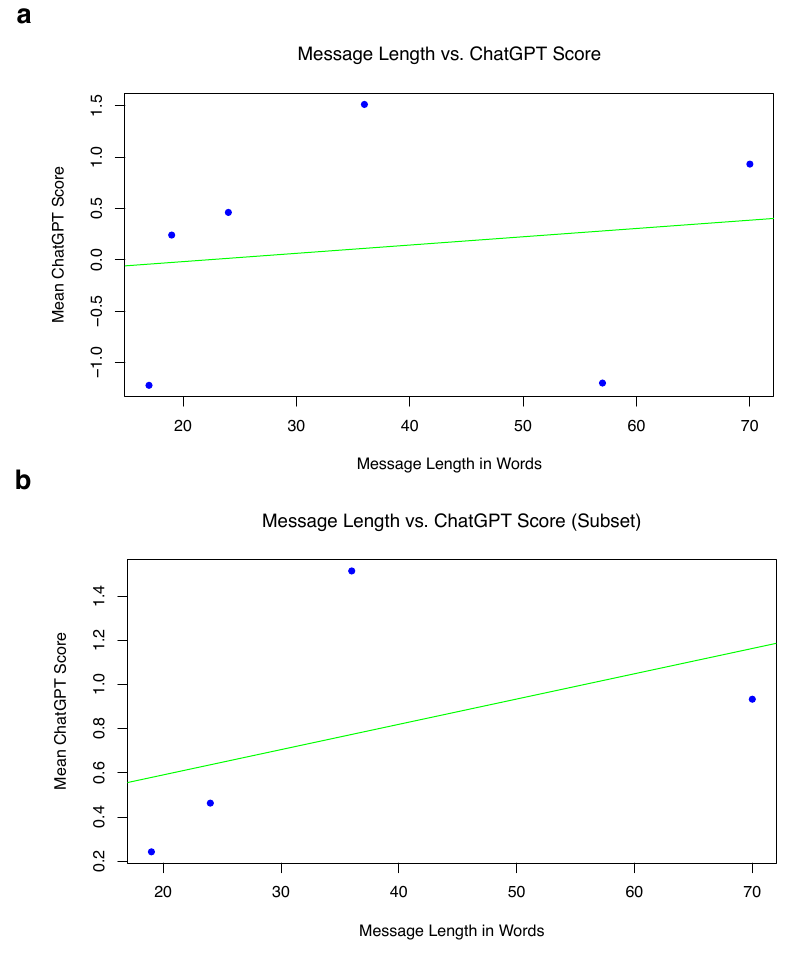}
\caption{Longer ChatGPT-augmented messages are more likely to be rated higher than shorter ones. We plot a message's length and its subjects' evaluation score for all six ChatGPT-augmented messages ($n=138$, correlation coeff $r = 0.00807$, intercept =-0.17740  p-value 0.7661 ) in panel (a) and in panel (b) we focus on the four ChatGPT-augmented messages that were placed first ($n=138$, correlation coeff $r = 0.01145$, intercept =0.36207  p-value  0.5332).}
\label{fig3}
\end{figure}

Interestingly, we observed a positive relationship between the length of ChatGPT-augmented message and its subjects' evaluation score. As shown in Fig.~\ref{fig3}a, simple linear modeling yielded an estimated correlation coefficient of 0.00807, suggesting a small positive relationship between message length and score. In this case, the p-value is 0.7661, suggesting that the model as a whole is not statistically significant at the conventional significance level of 0.05. Further subsetting the messages where the ChatGPT-augmented message was listed first, we found that the same linear model became slightly improved after this subset was taken (Fig.~\ref{fig3}b). The correlation coefficient slightly increases to 0.01145, again suggesting a small positive relationship between message length and score. The p-value for this model was lower than the first model, but still not statistically significant with a p-value of 0.5332 exceeding the conventional significance level of 0.05. While lacking sufficient statistical power, these results suggest that one factor that could be examined in future research on ChatGPT-augmented messaging is a message's length. ChatGPT may have more room for editing longer messages, thereby being better able to augmenting them~\cite{wang2023survey}. 


\begin{figure}[htp]
\includegraphics[width=13cm]{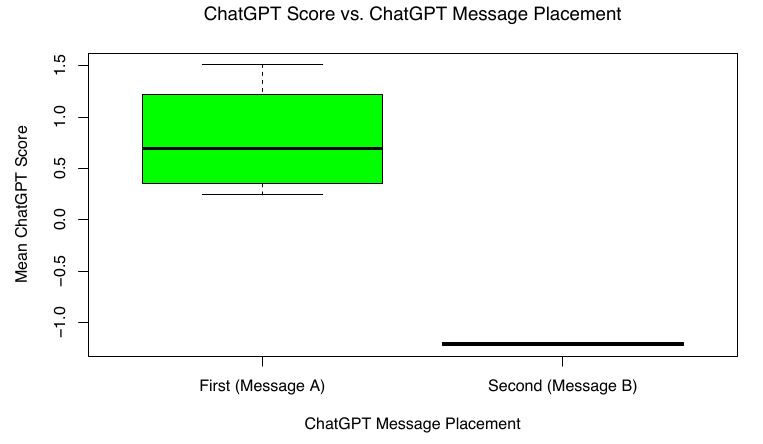}
\caption{Sequential effect of message placement order. While ChatGPT-augmented messages are placed randomly, either before or after the original, in our A/B testing, subjects evaluated the ones placed first significantly higher than those placed second (mean first: 0.788603, second:: -1.209558). The sample size was $n = 138$, and the Welch two-sample t-test, with an alternative hypothesis that the true difference in means is not equal to zero, yielded a p-value of 0.005718, which at an alpha level of 0.05, is statistically significant.}
\label{fig4}
\end{figure}

Apart from the scope and content of messages, there was a strong relationship between a message's placement in the survey question and how it was rated. This relationship can offer insights into much of the differences in the ChatGPT score variation, as shown in the boxplots in Fig.~\ref{fig4}. In the four questions where the ChatGPT-augmented message was positioned first, its mean score was positive at $0.788603$, whereas in the remaining two questions where the ChatGPT message was positioned second, its mean score was negative at $-1.209558$. 

A potential explanation for this result is that people are prone to selecting the first message they see, either because psychologically that makes a difference or because they are not responding properly to the survey.
Given the randomization we applied to all A/B testing questions, it is important to note this sequential effect of message placement reflects the primacy effect ---placing an option first often leads to a preference bias, as studied in previous research~\cite{murphy2006primacy}. This underscores the importance of message placement order in surveys, and accounting for the trade-eoff between primacy effect (options being placed first) and recency effect (options being placed second) would potentially lead to different methodology being used in future surveys.

\section{Discussion and Conclusion}

To shed light on our first finding---that ChatGPT messages are generally rated higher than original messages---we examine the characteristics of ChatGPT messages. In particular, four of our ChatGPT-augmented messages were rated significantly higher than the original messages. We want to explain potential reasons for the variance in message ratings and what factors contribute to ChatGPT being more effective as message augmentation under certain circumstances. 

The ChatGPT-augmented messages in our study were generally more enthusiastic and contained more exclamation points than the original messages. This could be attributed to what ChatGPT interprets an interesting and thus effective message to be. ChatGPT-augmented messages also tended to display more urgency than the original messages, including phrases such as ``Don't wait!'' or ``Get vaccinated today!'' that were not present in the original messages. A potential explanation for this pattern is that ChatGPT interprets effective messages as those containing calls for immediate action in order to  persuade~\cite{cialdini2001science}.

Given that our survey results pointed in favor of ChatGPT but not decisively so, we are led to recommend an iterative approach for agencies or individuals creating and refining messaging using ChatGPT collaboratively~\cite{chen2024large}. We envision a potential synergy between humans and ChatGPT~\cite{vaccaro2024combinations}. Our results showed a positive correlation between ChatGPT score and message length in word count, so we see this synergy being particularly effective with longer messages. Additionally, our results do not indicate that parties interested in leveraging ChatGPT for augmenting messaging purposes should not be too concerned about the public perception of using ChatGPT. Our study contributes to the emerging literature on studying persuasion with LLMs in a variety of domains for a good society~\cite{rescala2024can,rogiers2024persuasion,salvi2024conversational}, from increasing vaccination~\cite{sohail2024chatgpt} to improving health messaging~\cite{lim2023artificial}.

Our study has some limitations that should be discussed. First, future iterations of this study might randomize the order of A/B testing, particularly whether the ChatGPT-augmented messages appear before or after the original ones each time, to minimize the effects of message placement order on responses. Meanwhile, a larger sample size is needed for further studies to establish statistically significant relationships between message source (original vs ChatGPT), message score, and message length.

Second, the messages in our study were all targeted to a general audience, but ChatGPT could be effective at tailoring messages to individuals based on their demographics or specific concerns related to vaccination.  A recent study demonstrates that such personalized messaging bots based on LLMs are effective in reducing conspiracy beliefs~\cite{costello2024durably}. A Similar method can be used to enhance the public's confidence in vaccination.    

Finally, future research could experiment further with more sophisticated prompt engineering methods~\cite{cox2023prompting}, possibly using different LLMs such as LLAMA and GPT-4o, and have respondents rank them, rather than researchers making a judgment call on which prompts yield the best results.

Last but not least, future applications of this research may involve changing the messaging topic to one related to voting~\cite{potter2024hidden} or political campaigns~\cite{hackenburg2024evaluating}. To gather more diverse data from a range of message sources, another MTurk survey could involve asking participants to find a message, input it into ChatGPT, and compare the two messages, although this crowdsourcing method would eliminate the blind A/B testing aspect of our study.

\section{Data and Methods}

\paragraph{Overall methods.}
ChatGPT generated vaccination messages are generally very formulaic and unoriginal. For example:

\begin{quote}
  ``Getting vaccinated against COVID-19 is a crucial step to protect yourself and those around you. It's a simple yet powerful way to contribute to the collective effort in overcoming this global challenge. Encourage others to get vaccinated too and let's work together for a safer, healthier future.''\\  
``Vaccination plays a key role in controlling the spread of COVID-19. By getting vaccinated, you not only safeguard your own health but also contribute to the well-being of the community. Let's prioritize public health and encourage everyone to take this important step in the fight against the pandemic.''
\end{quote}

\noindent
For that reason, we focused on ChatGPT's capability to augment rather than create messages.

\paragraph{Message selection.}
Presumably human-generated messages were gathered from a variety of web sources. For the purpose of testing the ChatGPT message augmentation method in a range of use cases, sources varied from small, local public health departments to large companies. Media types mostly consisted of Tweets of varying lengths (six messages in total), as well as a transcription of a video ad for a pneumococcal vaccine. This seventh message was included only for piloting purposes and was excluded from our analyses. Specific sources included the Center for Disease Control, the World Health Organization, a local public health department, the National Health Service, and one individual Twitter user. 

\paragraph{Prompt engineering.}
Based on preliminary testing of different prompts given to ChatGPT, the prompt that yielded the most change to the original messages was "Make this message more interesting [insert message]". This prompt often yielded messages that were much longer than the original messages. For that reason, we modified the messages by adding a word limit at the end of the message. The final prompt was of the form "Make this message more interesting in [original message word count + 5] words [insert message]."

\paragraph{Survey design.}
 The main survey consisted of four demographic questions, seven message comparison questions, one comprehension test, and two ChatGPT questions. The survey was described as a study on vaccination messaging, with no mention of ChatGPT in the title or description. The four demographic questions asked for age, race, gender, and highest level of education achieved. The six message comparison questions each listed two messages, labelled A and B, one of which was the original message and one of which was that message augmented by ChatGPT. 
 
 Participants were asked to select which message they found more convincing, and then rate from 1-5 how much more convincing they found it compared to the previous message. The comprehension test asked the participants a question that required them to select a specific response. The final part of the survey revealed to respondents that half the messages in the survey were augmented by ChatGPT. Respondents were asked how knowing a message was generated by ChatGPT would influence their opinion of the message on a sliding scale from -3 to 3. Respondents were also given an option to provide a comment elaborating on their views of ChatGPT augmented messaging. 

\paragraph{Survey administration.}
The main survey was released on Amazon Mechanical Turk for 200 respondents. Amazon Mechanical Turk, or MTurk, is a platform that allows people to hire remote workers to complete tasks. The survey was described as a study on vaccination messaging. Participants were given a link to the Qualtrics Survey and compensated with 10 cents upon completion of the survey. 
Several methods were used to improve survey reliability. First, the "Prevent multiple responses" feature on Qualtrics was enabled and responses with the same IP address were flagged and removed from results. Also, the survey asked a question that served as a comprehension test that required respondents to select a specific answer. Any respondents who selected an incorrect answer had their other responses removed. We used results from the Qualtrics duplicate score and robot score to remove more responses. Of the 200 original responses, 136 were used in our final results.

\paragraph{Survey analysis.}

In order to analyze our results, whenever respondents indicated they preferred a ChatGPT-augmented message, we assigned that response a score of 1 and if they preferred the original message, we assigned it a score of -1. To calculate a message’s overall rating, weighted by the degree to which the respondent preferred that message, we multiplied the weighting factor by 1 or -1 to create an overall score ranging from -5 to 5. This data was used to create Figure 1.
We then calculated the mean ChatGPT score for each respondent and compared it with their response on how positively or negatively they viewed ChatGPT using a Pearson Correlation test. 

We also tested how a message’s placement in the survey impacted its ratings (i.e. whether the message was listed as message A or B). We constructed two boxplots that showed the message’s placement compared to its average ratings and visually inspected it. We then conducted a Welch two sample t-test to determine whether the difference in means was statistically significant. 
After noticing a correlation between a message’s placement and its scores, we conducted a partial correlation analysis that controlled for a message’s placement while looking at a message’s score and comparing it with its length.
Additionally, we performed ANOVA testing with several demographic variables (education, age, race) against mean ChatGPT scores and found no results of statistical significance. We also performed a t-test of the gender variable compared with mean ChatGPT scores and found no statistically significant relationship.

\section*{Acknowledgements.}
Supported in part by a research grant from Investigator-Initiated Studies Program of Merck Sharp \& Dohme Corp. The opinions expressed in this paper are those of the authors and do not necessarily represent those of Merck Sharp \& Dohme Corp.

\end{document}